\documentclass[11pt]{revtex4-2}

\usepackage{hyperref}
\usepackage{braket}
\usepackage{amsmath,amsfonts,amssymb}
\hypersetup{dvips,dvipdfm,colorlinks=true,urlcolor=magenta,filecolor=magenta,linktoc=page,citecolor=red,linkcolor=blue,bookmarks=true}
\usepackage{graphicx,epstopdf}

\begin{document}
	\title{\textbf{ Classical and Quantum Cosmology in Einstein-\ae ther Scalar-tensor gravity: Noether Symmetry Analysis}}
	\author{ Dipankar Laya$^1$\footnote {dipankarlaya@gmail.com}Roshni Bhaumik$^1$\footnote {roshnibhaumik1995@gmail.com} Sourav Dutta$^2$\footnote {sduttaju@gmail.com} Subenoy Chakraborty$^1$\footnote {schakraborty.math@gmail.com}}
	\affiliation{$^1$Department of Mathematics, Jadavpur University, Kolkata-700032, West Bengal, India\\$^2$Department of Mathematics, Dr. Meghnad Saha College, Itahar, Uttar Dinajpur-733128, West Bengal, India.}

	\begin{abstract}
		The present work deals with Einstein--\ae ther Scalar tensor gravity	in the background of homogeneous and isotropic flat FLRW space--time model. The Noether symmetry vector identifies a transformation in the augmented space so that the field equations become solvable. The cosmological solutions are analyzed from the observational point of view. Finally, for quantum cosmology, the Wheeler--DeWitt (WD) has been formulated and solutions have been determined by identifying the periodic nature of the wave function using conserved (Noether) charge.
	\end{abstract}
	\maketitle

	\section{Introduction}
	A generally covariant modification of Einstein gravity is the Einstein--\ae ther theory \cite{r11, r12, r13, r14}. In this theory, in addition to the space--time metric there is a scalar field (named \ae ther) indicating a Universal notion of time. Thus the theory has a preferred frame of reference and as a result, there is a Lorentz violation of the theory (i.e., the theory breaks the invariance under boosts)\cite{r1, r2, r4, r5, r6, r7, r8, r9, r10}. Further, due to this symmetry breaking, a Higgs mechanism for the gravitation gives a variation at large distance physics and it may be a possible explanation for the recent Supernova data which is well explained by cosmological constant. One may note that Einstein--\ae ther theory goes over to Einstein gravity if it preserves locality as well as covariance formulation\cite{r14.1, r15, r16, r17, r18, r19, r20, r21}. Also, in the classical limit the Horava-Lifshitz\cite{r22} gravity corresponds to this modified gravity theory but not in reverse way\cite{r23, r24}. Moreover, this Lorentz violating gravity theory is important from the point of view of quantum gravity theory. In recent past, gravitational models with Lorentz violation are of special interest in the literature.
	
	Scalar filed in cosmology plays an important role for the description of the evolution of the Universe. Such models start with Brans-Dicke theory of gravity. This gravity theory obeys Mach's principle\cite{r25} and is described by the space--time metric with non--minimally coupled scalar field. This scalar field plays an important role to identify the early era of evolution\cite{r26}. Further, the higher--order derivatives of the scalar field in the gravitational action integral indicates quantum corrections or modifications of Einstein gravity. In Einstein--\ae ther theory if the scalar field is non--minimally coupled then it may be termed as Einstein--\ae ther scalar tensor theory\cite{r27, r28, r29}. There is another well known scalar field model known as dilation filed which is used for description of string cosmology\cite{r30}.\\
	
	{In this context it is worthy to mention that there is monograph on Noether symmetries in theories of Gravity \cite{Bajardi:2022ypn}. In this monograph various alternatives  and extensions to Einstein gravity has been examined form symmetry principles to identify physically viable models. Subsequently, the authors constrains the parameters of the viable models by Noether symmetry analysis and compare them with recent observational data.}   
	
	The present work is an attempt to study the cosmic evolution in Einstein--\ae ther scalar--tensor theory with background FLRW space--time model both from classical and quantum cosmological view point. Due to homogeneity and isotropy of the space--time, the E\ae ther field is a comoving time--like vector field and the field equations are second order coupled nonlinear differential equations. The well known Noether symmetry has been imposed both for determining the unknown functions parameters in the system and to identify a transformation in the augmented space so that one of the variable becomes cyclic {(for application in scalar tensor gravity one may see the works in references \cite{mu1,mu2,mu3,mu4})}. As a results the field equations get simplified and become solvable to have classical solutions. {Noether symmetry analysis has an important role in describing quantum cosmology (for application of Noether symmetry in quantum cosmology one may see the works in references \cite{mu5,mu6,mu7,mu8,mu9})} Noether symmetry identifies the oscillatory part of the wave function of the Universe by writing the conserved (Noether) charge equation in operator form. As a result the Wheeler-Dewitt (WD) equation becomes simplified and is solvable using separation of variables. The probability measure identifies whether initial singularity can be avoided or not by quantum description. The plan of the paper is the following:\\
	An overview of this model has been presented in Section {\bf II}. In Section {\bf III} we have presented a basic concept of Noether symmetry analysis and in Section {\bf IV} we are able to find the analytical solution of this model using Noether symmetry analysis. Physical metric and its symmetry analysis are presented in Section {\bf V}. Section {\bf VI} deals with the analysis of quantum cosmology of this model and finally, we draw our conclusions in Section {\bf VII}.
	
	\section{Einstein--\ae ther scalar tensor gravity: an overview}
	
	The Action integral for the present model can be written as
	\begin{equation}
		S=S_1+S_2,\label{e1}
	\end{equation} 
	where $S_1$ stand for the action for the scalar-tensor theory and $S_2$ represents action for the Einstein--\ae ther theory. In Jordan frame scalar--tensor action takes the form:
	\begin{equation}
		S_1=\int d^4 x \sqrt{-g}\bigg(A(\phi)R+\frac{1}{2}g^{\mu \nu}\phi_{, \mu} \phi_{, \nu}+V(\phi)\bigg),\label{e2}
	\end{equation}
	where $A(\phi)$ stand for the nonminimal coupling of the scalar field $\phi$ with gravity and $V(\phi)$ is the potential function of the scalar field. If $\sigma^{\mu}$ denotes the \ae ther field then corresponding action integral takes the form
	\begin{eqnarray}
		S_2=-\int d^4 x \sqrt{-g}\bigg[\alpha_1(\phi)&\sigma^{\nu;\mu}& \sigma_{\nu;\mu}+\alpha_2(\phi) (g^{\mu \nu}\sigma_{\mu;\nu})^2+\alpha_3(\phi)\sigma^{\nu;\mu}\sigma_{\mu;\nu}\nonumber\\
		&+&\alpha_4(\phi)\sigma^{\mu} \sigma^{\nu} \sigma_{;\mu}\sigma_{\nu}-\delta(\sigma^{\mu}\sigma_{\nu}+1)\bigg],\label{e3}
	\end{eqnarray}
	where $\alpha_i$'s ($i=1,2,3,4)$ are the coefficient function identifying the coupling between the \ae ther field and the scalar field, te Lagrange multiplier `$\delta$' indicates the unitarity of the \ae ther field i.e., $\sigma^{\mu}\sigma_{\mu}=-1$. Now due to cosmological principle the space--time geometry is described by homogeneous and isotropic FLRW line element. As a result, the explicit form of the action integral takes the form
	$$S=\int L dt$$
	with the point like Lagrangian \cite{r31} 
	\begin{equation}
		L=6\mu(\phi)a \dot{\phi}^2+6 \xi(\phi)a^2\dot{a}\dot{\phi}+\frac{1}{2}a^3\dot{\phi}^2-a^3V(\phi).\label{e4}
	\end{equation}
	Here $\mu(\phi)=A(\phi)+\frac{1}{2}\big(\alpha_1(\phi)+3\alpha_2(\phi)+\alpha_3(\phi)\big)$ and $\xi(\phi)=\frac{dA}{d\phi}$ and `$a$' is the scale factor of the Universe. It is to be noted that this minisuperspace description is very much suitable for gravitational models. In fact, the field equations corresponding to the above point Lagrangian identify the evolution of a point--like particle while minisuperspace is very much essential for canonical quantization and as a result, it is possible to formulate the WD equation in quantum cosmology. Now varying the above Lagrangian (\ref{e4}) with respect to the variable $a$ and $\phi$ the field equations are
	\begin{equation}
		2\dot{H}+3H^2+2\frac{\mu_{,\phi}}{\mu}H\dot{\phi}-\frac{\bigg(\frac{\dot{\phi}^2}{2}-V(\phi)\bigg)}{2}+\frac{\xi}{\mu}\bigg(\frac{\xi_{,\phi}}{\xi}\dot{\phi}^2+\ddot{\phi}\bigg)=0,\label{e5}
	\end{equation}
	\begin{equation}
		\ddot{\phi}-\dot{\phi}+3H\dot{\phi}+6\xi(\phi)(\dot{H}+3H^2)-6\mu(\phi)_{,\phi}~ H^2+V_{,\phi}=0,\label{e6}
	\end{equation}
	and also we have the scalar constraint equation
	\begin{equation}
		6\mu(\phi)~H^2+6\xi(\phi)~H\dot{\phi}+\frac{\dot{\phi}^2}{2}+V(\phi)=0,\label{e7}
	\end{equation}
	where $H=\frac{\dot{a}}{a}$ is the usual Hubble parameter. Now the field equations (\ref{e5}) and (\ref{e7}) can be rewritten in the form of the usual Friedmann equations as
	\begin{equation}
		3H^2=G_{\mbox{mod}}~\rho_{\mbox{mod}},\label{e8}
	\end{equation}
	and
	\begin{equation}
		2\dot{H}+3H^2=G_{\mbox{mod}}~p_{\mbox{mod}},\label{e9}
	\end{equation}
	where $G_{\mbox{mod}}=\frac{1}{-\mu(\phi)}$ is the modified time varying gravitational constant and the modified energy density and the thermodynamic pressure have the expressions
	\begin{equation}
		\rho_{\mbox{mod}}=\frac{1}{2}\bigg(6\xi(\phi)~H\dot{\phi}+\frac{\dot{\phi}^2}{2}+V(\phi)\bigg),\label{e10}
	\end{equation}
	\begin{equation}
		p_{\mbox{mod}}=-\bigg[-2\frac{d\mu}{d\phi}~H\dot{\phi}-\frac{1}{2}\bigg(\frac{\dot{\phi}^2}{2}-V(\phi)\bigg)+\frac{d\xi}{d\phi}\dot{\phi}^2+\xi(\phi)~\ddot{\phi}\bigg].\label{e11}
	\end{equation}
	Note that gravitational coupling parameter depends on the coupling function $A(\phi)$ and the \ae ther coefficients $\alpha_i (i=1,2,3,4)$. For constant $\alpha_i$'s the above model reduces to usual scalar--tensor theory.
	\section{An overview of Noether symmetry analysis and classical solutions}
	According to Noether, if the Lagrangian of physical system is invariant with respect to the Lie derivative along a vector field in the augmented space (i.e.,$~	\mathcal{L}_\mathcal{\overrightarrow{\chi}}L=\overrightarrow{\chi}L=0$) then the physical system must be associated to some conserved quantities (Noether 1st theorem)\cite{r32, r33, r34, r35, r36}, known as Noether charge. In addition, the symmetry constraints lead the evolution equations to be solvable or the evolution equations in a much simplified form.
	
	In general, for a point-like canonical Lagrangian $L=L[q^\alpha(x^i),\partial_jq^\alpha(x^i)],~q^{\alpha}(x^i)$ being the generalized coordinates, the Euler-Lagrange equations are
	\begin{equation}
		\partial_j{\left( \frac{\partial L}{\partial(\partial_j q^\alpha)}\right)}-\frac{\partial L}{\partial q^\alpha}=0~,~\alpha=1,2,...N\label{e12}
	\end{equation}
	with $N$, the dimension of the augmented space. Now contracting the above Euler-Lagrange equations with some unknown function $\mu^\alpha(q^\beta)$ and simplifying algebraically one gets 
	\begin{equation}
		\mathcal{L}_\mathcal{\overrightarrow{\chi}}L=
		\mathcal{\overrightarrow{\chi}}L=\lambda^\alpha\frac{\partial L}{\partial q^\alpha}+ (\partial_j\lambda^\alpha)\left(\frac{\partial L}{\partial(\partial_jq^\alpha)}\right)=\partial_j\left(\lambda^\alpha\frac{\partial L}{\partial(\partial_j q^\alpha)}\right)\label{e13}
	\end{equation}
	where
	\begin{equation}
		\overrightarrow{\chi}=\lambda^\alpha\frac{\partial}{\partial q^\alpha}+(\partial_j\lambda^\alpha)\frac{\partial}{\partial(\partial_jq^\alpha)}\label{e14}
	\end{equation}
	is the vector field in the augmented space (i.e., the space consists of the generalized coordinates and their derivatives). Now according to Noether's theorem if $~\mathcal{L}_\mathcal{\overrightarrow{\chi}}L=0$, then
	$\mathcal{\overrightarrow{\chi}}$ (in equation (\ref{e14})) is termed as Noether symmetry vector field or infinitesimal generator of the Noether symmetry and the associated conserved Noether current has the expression 
	\begin{equation}
		I^i=\lambda^\alpha\frac{\partial L}{\partial(\partial_i q^\alpha)}
	\end{equation}\label{e15}
	with $\partial_iI^i=0$.
	
	In the present problem $N=3$ i.e., the 3D augmented space consists of the variables $(t,a,\phi)$. Further, due to explicit time independence of the Lagrangian the Hamiltonian i.e., the total energy function is a constant of motion for the system i.e.,
	\begin{equation}
		E=\dot{q}^\alpha\frac{\partial L}{\partial\dot{q}^\alpha}-L=\mbox{conserved} \label{e16}
	\end{equation} 
	In addition, if the conserved quantity due to the symmetry has some physical analog then this symmetry approach can identify the reliable model.
	
	On the other hand, Hamiltonian formulation is more useful than Lagrangian description in the context of quantum Cosmology. Then the Noether's theorem can be stated as 
	\begin{equation}
		\mathcal{L}_{\overrightarrow{X}_H}H=0\label{e17}
	\end{equation}
	with $~\overrightarrow{X}_H=\dot{q}\frac{\partial}{\partial q}+\ddot{q}\frac{\partial}{\partial \dot{q}}$ as the symmetry vector. Moreover, in quantum cosmology Hamiltonian formulation is more useful than Lagrangian formalism. One may note that the canonically conjugate momenta corresponding to the conserved current due to Noether symmetry is constant i.e., 
	\begin{equation}
		\pi_{\rho}=\frac{\partial L}{\partial q^{\rho}} =p_{0\rho}~,~~~\mbox{a constant}
	\end{equation}\label{e18}
	with $\rho=1,2,...m$ indicating the no of symmetries. So in quantum formulation by writing the operator version of equation (18) one gets
	\begin{equation}
		-i\partial_{q^\rho}|\psi>=p_{0\rho}|\psi>\label{e19}
	\end{equation}
	where $|\psi>$ is termed as the wave function of the universe. Thus solving equation (\ref{e19}) one gets the oscillatory part of the wave function. So the wave function of the universe has the explicit form as
	\begin{equation}
		|\psi>=\sum_{\rho=1}^{m}e^{ip_{0\rho}q^{\rho}}|\phi(q^\sigma)>,~~\sigma<N\label{e20}
	\end{equation}
	Here `$\sigma$' indicates a direction along which there is no symmetry and $N$ is the dimension of the minisuperspace. The oscillatory part of the wave function indicates Noether symmetry along that direction and the corresponding conjugate momenta is conserved and vice-versa. In other words the symmetry vector identifies the cyclic variables in the physical system.
	\section{Noether symmetry analysis and classical solution to the present modified gravity theory}
	This section shows how the above coupled non-linear field equations (\ref{e5})-(\ref{e7}) can be solved as an application of the Noether symmetry analysis. According to this symmetry, $\exists$ a function $G(t,a,\phi)$ such that the Lagrangian satisfies the equation
	\begin{equation}
		X^{[1]}L+LD_t\eta(t,a,\phi)=D_tG(t,a,\phi)\label{e21}
	\end{equation}
	with symmetry vector
	\begin{equation}
		X=\eta(t,a,\phi)\frac{\partial}{\partial t} + 
		\alpha(t,a,\phi)\frac{\partial}{\partial a} + 
		\beta(t,a,\phi)\frac{\partial}{\partial\phi}\label{e22}
	\end{equation}
	Here the total derivative operator $D_t$ is given by 
	\begin{equation}
		D_t\equiv\frac{\partial}{\partial t}+\dot{a} \frac{\partial}{\partial a}+\dot{\phi}\frac{\partial}{\partial\phi}\label{e23}
	\end{equation}
	with $X^{[1]},~$ the first prolongation vector as 
	\begin{equation}
		X^{[1]}=X+(D_t\alpha-\dot{a}D_t\eta)\frac{\partial}{\partial\dot{a}}+
		(D_t\beta-\dot{\phi}D_t\eta)\frac{\partial}{\partial\dot{\phi}}\label{e24}
	\end{equation}
	Then the conserved quantity associated with the vector field $X$ has the expression 
	\begin{equation}
		I=\eta L+(\alpha-\dot{a}\eta)\frac{\partial L}{\partial  \dot{a}}+
		(\beta-\dot{\phi}\eta)\frac{\partial L}{\partial \dot{\phi}}-G\label{e25}
	\end{equation}
	For the Noether symmetry of the Lagrangian(\ref{e4}) on the tangent space $(a,\dot{a},\phi,\dot{\phi})$, the constituent partial differential equations are 
	\begin{eqnarray}
		\eta_a=\eta_{\phi}&=&0\nonumber\\
		-3\alpha a^2 V(\phi)-a^3\beta V'(\phi)-a^3\eta_t V(\phi)&=&G_t\nonumber\\
		12\mu(\phi)a\alpha_t+6\xi(\phi)a^2\beta_t&=&G_a\nonumber\\
		6\xi(\phi)a^2\alpha_t+a^3\beta_t&=&G_{\phi}\nonumber\\
		6\mu(\phi)\alpha+6a\mu'(\phi)\beta+12a\mu(\phi)\alpha_a-12\mu(\phi)a\xi_t+
		6\xi(\phi)a^2\beta_a+6\mu(\phi)a\eta_t&=&0\nonumber\\
		\frac{3}{2}\alpha a^2+6\xi(\phi)a^2\alpha_{\phi}-a^3\eta_t+a^3\beta{\phi}
		+\frac{a^3\eta_t}{2}&=&0\nonumber\\
		12\xi(\phi)a\alpha+6\xi'(\phi)a^2\beta+6\xi(\phi)a^2\alpha_a+12\mu(\phi)a\alpha_{\phi}-
		\nonumber\\6\xi(\phi)a^2\eta_t+a^3\beta_a+6\xi(\phi)a^2\beta_{\phi}&=&0\nonumber
	\end{eqnarray}
	Thus due to Noether symmetry the coefficients (i.e., $\eta,\alpha,\beta$) of the infinitesimal generator satisfy an overdetermined set of partial differential equations which are solved using the method of separation of variables as 
	\begin{equation}
		\alpha=\alpha_0 a^{p_1}\phi^{q_1}~,~ \beta=\beta_0 a^{p_2}\phi^{q_2}
	\end{equation}\label{e26}
	Therefore we get the solution of above set of differential equation is given 
	below 
	\begin{eqnarray}
		\alpha=\alpha_0a^{p_1}\phi^{q_1}~~,~~\beta=k\alpha_0a^{p_1-1}\phi^{q_1+1}~~,
		~~V(\phi)=V_0\phi^{-\frac{3}{k}}~~,~~\nonumber\\\mu(\phi)=\frac{B_0}{2}\phi^2
		~~,~~\xi(\phi)=B_0\phi.\label{e26.1}
	\end{eqnarray}
	where $\alpha_0,B_0,V_0,k,$ being any constant. Hence the infinitesimal generator on the tangent space can be represented as
	\begin{equation}
		X^{[3]}=(\alpha=\alpha_0a^{p_1}\phi^{q_1})\partial_a+
		(\beta=k\alpha_0a^{p_1-1}\phi^{q_1+1})\partial_{\phi}
	\end{equation}\label{e26.2}
	
	
	\begin{figure}[h]
		\begin{minipage}{0.47\textwidth}
			\centering \includegraphics[height=5cm,width=8cm]{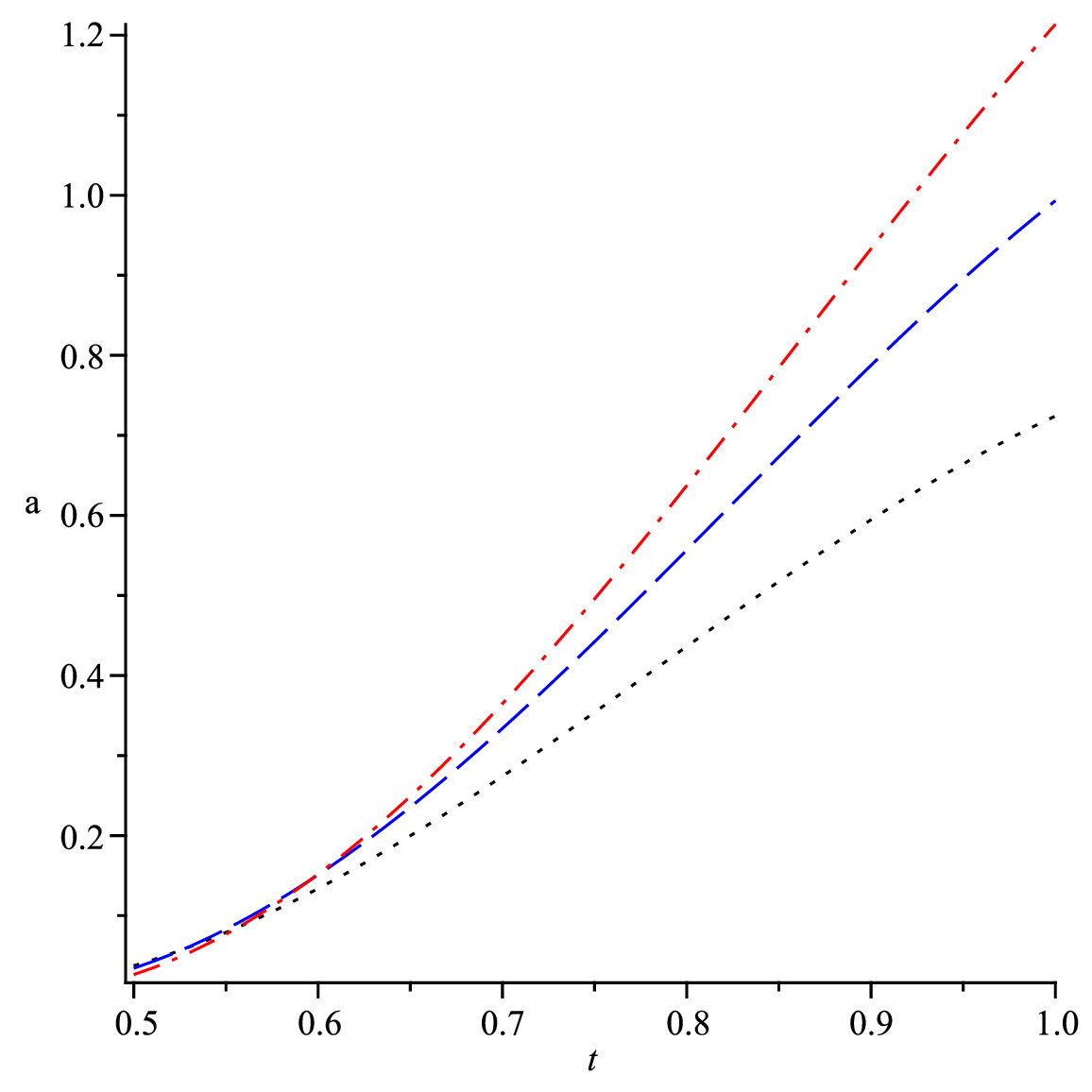}
		\end{minipage}\hfill
		\begin{minipage}{0.47\textwidth}
			\centering \includegraphics[height=5cm,width=8cm]{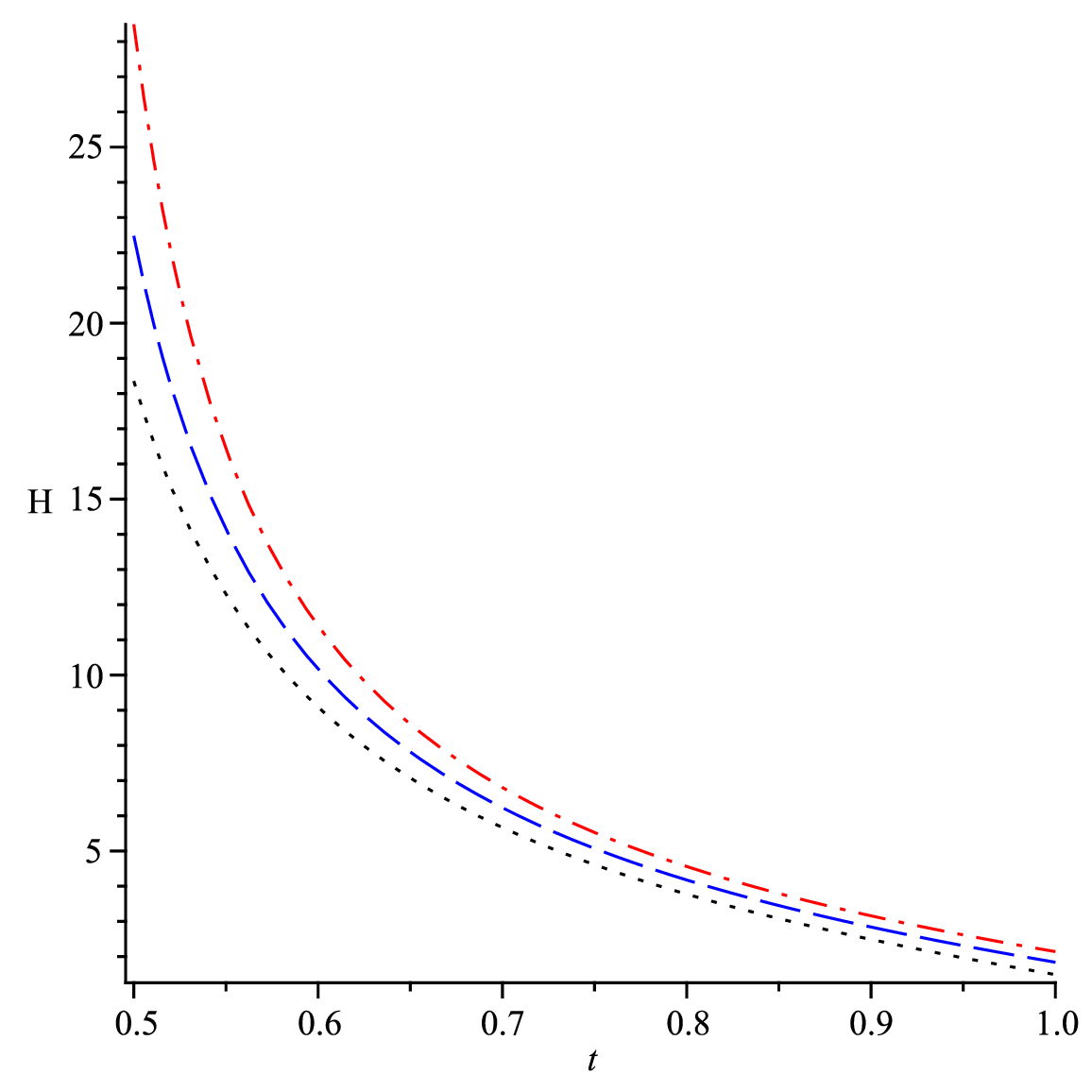}
		\end{minipage}
		\begin{minipage}{0.47\textwidth}
			\centering \includegraphics[height=5cm,width=8cm]{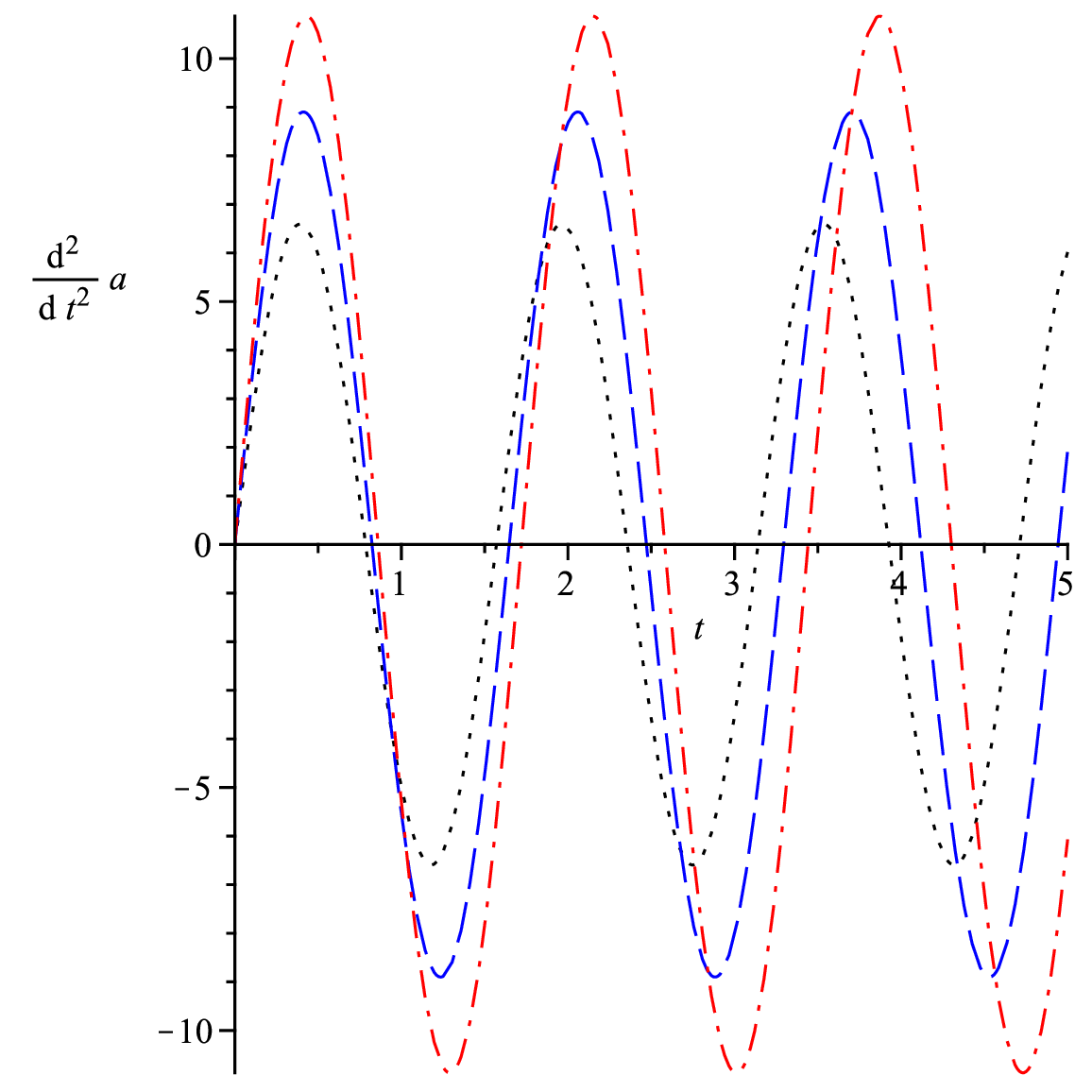}
		\end{minipage}
		\caption{The graphical representation of scale factor $a(t)$ (top left), Hubble parameter $H(t)$ (top right) and acceleration parameter $\ddot{a}(t)$ (bottom) with respect to cosmic time $t$.}\label{f1}
	\end{figure}
	
	Thus symmetry analysis not only gives the symmetry
	vector but also determine the potential function and the coupling function. Further, the explicit form of the conserved current and conserved energy are gives as 
	\begin{eqnarray}
		I^{\alpha}&=&\alpha\frac{\partial L}{\partial\dot{a}}+\beta\frac{\partial L}{\partial 
			\dot{\phi}}\\\label{e27}
		\mbox{and}~~E&=&\dot{a}\frac{\partial L}{\partial\dot{a}}+\dot{\phi}\frac{\partial 
			L}{\partial \dot{\phi}}-L\label{e28}
	\end{eqnarray}   
	Now, in order to identify a point transformation $(a,\phi)\rightarrow(u,v)$ so that one of the variables becomes cyclic we impose the restriction \cite{r37} 
	\begin{equation}
		i_Xdu=1 ~\mbox{and}~~ i_Xdv=0\label{e29}
	\end{equation}
	Then the transformed infinitesimal generator 
	\begin{equation}
		\hat{X}=(i_Xdu)\frac{\partial}{\partial 
			u}+(i_Xdv)\frac{\partial}{\partial v}
		+\left(\frac{d}{dt}(i_Xdu)\right)\frac{d}{d\dot{u}}+\left(\frac{d}{dt}(i_Xdv)\right)
		\frac{d}{d\dot{v}}\nonumber
	\end{equation} 
	simplifies to  $\hat{X}=\frac{\partial}{\partial u}$ with $\frac{\partial 
		L}{\partial u}=0$ i.e., $u$ becomes a cyclic variable. Now solving the equation (\ref{e29}) one can get the relation between the old and new variable as
	\begin{eqnarray}
		u&=&-\frac{1}{3}\ln\Big(a^{-\frac{3}{2}}\phi\Big)\\\label{e32}
		v&=&\ln\Big(\frac{a^\frac{3}{2}}{\phi}\Big)\label{e33}
	\end{eqnarray}
	As a consequence the transformed Lagrangian takes the form 
	\begin{equation}
		L=e^{-2v}[A_1\dot{u}^2+B_1\dot{v}^2+C_1\dot{u}\dot{v}-V_0]\label{e34}
	\end{equation}
	
	with the conserved energy given by
	\begin{equation}
		E=e^{-2v}[A_1\dot{u}^2+B_1\dot{v}^2+C_1\dot{u}\dot{v}+V_0].\label{e35}
	\end{equation}
	Then by forming the Euler-Lagrange equation from equation (\ref{e34}), we get the new variable as
	\begin{eqnarray}
		u(t)&=&-\frac{C_1}{4A_1}\ln\Big[\frac{m}{2V_0}\Big(C_0\sin\Big(\sqrt{\frac{2V_0}{m}t}\Big)-
		C_2\cos\Big(\sqrt{\frac{2V_0}{m}}t\Big)\Big)^2\Big]+D\\\label{e36}
		v(t)&=&-\frac{1}{2}\ln\Big[\frac{m}{2V_0}\Big(C_0\sin\Big(\sqrt{\frac{2V_0}{m}t}\Big)-
		C_2\cos\Big(\sqrt{\frac{2V_0}{m}}t\Big)\Big)^2\Big]\label{e37}
	\end{eqnarray}
	Thus the classical cosmological Solution of this Einstein-\ae ther field model are
	\begin{eqnarray}
		a(t)=e^{-\big(\frac{2A_1+C_1}{4A_1}\big)\ln\left\{\frac{m}{2V_0}\Big(C_0\sin\Big(\sqrt{\frac{2V_0}{m}t}\Big)-
			C_2\cos\Big(\sqrt{\frac{2V_0}{m}}t\Big)\Big)^2\right\}+D}\\\label{e38}
		\phi(t)=e^{-\big(\frac{2A_1-C_1}{4A_1}\big)\ln\left\{\frac{m}{2V_0}\Big(C_0\sin\Big(\sqrt{\frac{2V_0}{m}t}\Big)-
			C_2\cos\Big(\sqrt{\frac{2V_0}{m}}t\Big)\Big)^2\right\}-\frac{D}{2}}\label{e39}
	\end{eqnarray}
	The above cosmological solution has been presented graphically in FIG.\ref{f1}. The scale factor, Hubble parameter and the acceleration parameters has been plotted over cosmic time for different sets of parameter involved. From the figure one may conclude that the present model describes an expanding model of the Universe with Hubble parameter gradually decreases with the evolution. Also the Universe experiences an accelerated expansion in early era and subsequently there is a phase of decelerated expansion and finally, again the Universe is experiencing an accelerated expansion era of evolution. The nature of the cosmic evolution of the present model agrees with the observational evidences. Finally, as the cosmic parameters do not deviate significantly due to small changes in the parameter values so it is reasonable to consider the above cosmic solution to be stable in nature.
	
	\section{Physical metric and Symmetry analysis}
	
	It has been shown by Tsamparlis et.al \cite{r29}that if the Lagrangian of a physical system can be written in the form of $L=T-V$ then the Noether point symmetries are identified by the elements of the homothetic group of the kinetic metric. As the Lagrangian of the present problem (given by equation (\ref{e4})) can be expressed in the form of point particle so homothetic group of the kinetic metric will be very much relevant to identify the Noether point symmetries. The kinetic metric (obtained from the kinetic part of the Lagrangian) is given by
	\begin{equation}
		dS^2_{k}=6\mu(\phi)ada^2+6\xi(\phi)a^2dad\phi+\frac{1}{2}a^3(d\phi)^2
	\end{equation}
	with effective potential : $V_{\mbox{eff}}(a,\phi)=a^3 V(\phi)$.
	
	Substituting the values of $\mu(\phi)$ and $\xi(\phi)$  form equation (\ref{e26.1} ) and after a simple algebra one gets
	\begin{eqnarray}
		dS^2_k&=&\phi^2a^3\left[3B_0\left(\frac{da}{a}\right)^2+6B_0\frac{da}{a}\frac{d\phi}{\phi}+\left(\frac{d\phi}{\phi}\right)^2\right]\nonumber\\
		&=&e^{3\alpha+2u}\left[3B_0d\alpha^2+6B_0d\alpha du+\frac{1}{2}du^2\right]
	\end{eqnarray}
	with $e^\alpha=a, e^u=\phi$
	
	It is evident that the above metric is conformal to the metric 
	\begin{equation}
		dS^2_c=-d\alpha^2+d\beta^2\nonumber
	\end{equation}
	with $\beta=\frac{u+2\alpha}{\sqrt{2}}$ and $B_0=\frac{1}{3}$.
	
	Thus the given Lagrangian with the above transformation of variables simplifies to 
	\begin{equation}
		L=e^{3\alpha+2u}\left[-\left(\frac{\dot{a}}{a}\right)+\frac{1}{2}\dot{\beta} ^2+V(a,\beta)\right]
	\end{equation}
	Hence the present cosmological model can be consider to be equivalent to FLRW model (of scale factor `$a$') with a scalar field having coupled potential.
	
	Due to $2D$ con-formality of the kinematic metric, there exist the gradient Homothetic vector(HV) : $H_{\gamma}=a\partial_a$ with $e^{\beta}H_v=1$. Note that this HV does not generate a Noether symmetry of the Lagrangian. Further, there exists the Killing vectors of the $2D$  equation(\ref{e3}) and they span the $E2$ group. In addition, there are $4D$ homothetic Lie algebra with explicit form :
	
	(i) Two gradient translation Killing vectors :
	\begin{equation}
		\overrightarrow{X}^{(1)}=-\partial_{\alpha}~,~\overrightarrow{X}^{(2)}=\frac{1}{\alpha} \partial_{\beta}\nonumber
	\end{equation}
	having gradient Killing functions : $f^{(1)}_k=e^\beta~,~f^{(2)}_k=0$.
	
	(ii) one non-gradient Killing vector (indicating rotation) ;
	\begin{equation}
		\overrightarrow{X}^{(3)}=\partial_{\beta} \nonumber
	\end{equation}
	
	(iii) The gradient homothetic vector ;
	\begin{equation}
		\overrightarrow{X}^4=\alpha\partial_{\alpha}\nonumber
	\end{equation}

	\section{Quantum Cosmology in Einstein-\ae ther scalar-tensor gravity }
	In this present cosmological model, conjugate momenta associated to the $2$D configuration space $\{a,\phi\}$ can be written as
	\begin{eqnarray}
		p_a=\frac{\partial L}{\partial 
			\dot{a}}&=&6B_0a\phi(\phi\dot{a}+a\dot{\phi})\nonumber\\
		p_{\phi}=\frac{\partial L}{\partial 
			\dot{\phi}}&=&a^2(6B_0\phi\dot{a}+a\dot{\phi})\label{eqm1}
	\end{eqnarray}
	Then the Hamiltonian of the system takes the form as
	\begin{equation}
		H=\Big[\frac{A_2p_a^2}{a\phi}+\frac{B_2\phi p_{\phi}^2}{a^3}+\frac{C_2 
			p_ap_{\phi}}{a^2} -V_2a^3\phi^{1-\frac{3}{k}}\Big]\label{eqm2}
	\end{equation}
	with equivalent Hamilton's equation of motion
	
	\begin{eqnarray}
		\dot{a}&=&\left(\frac{6B_0\phi p_{\phi}-ap_a}{30B_0a^2\phi^2}\right)\nonumber\\
		\dot{\phi}&=&\left(\frac{ap_a-\phi p_{\phi}}{a^3\phi(6B_0-1)}\right)\nonumber\\
		\dot{p_a}&=&\left[\frac{A_2p_a^2}{a^2\phi}+\frac{3B_2\phi 
			p_{\phi}^2}{a^4}+\frac{2C_2 
			p_ap_{\phi}}{a^3} + 3V_2a^2\phi^{1-\frac{3}{k}}\right]\nonumber\\ 
		\dot{p_{\phi}}&=&\left[\frac{A_2p_a^2}{a\phi^2}-\frac{B_2 
			p_{\phi}^2}{a^3}+ V_2\Big(1-\frac{3}{k}\Big)a^3\phi^{\frac{3}{k}}\right]\label{eqm3}
	\end{eqnarray}
	
	In the context of Quantum cosmology, one has to construct the Wheeler Dewitt (WD) equation which is given by 
	\begin{equation}
		\hat{H}\psi(a,\phi)=0\label{eqm4}
	\end{equation}
	where $\hat{H}$ is the operator version of the Hamiltonian and $\psi(a,\phi)$ is the wave function of the Universe. In course of transformation to the operator version, there is a problem which is known as operator ordering problem. One has to consider the ordering consideration:
	$$p_a\rightarrow-i\frac{\partial}{\partial a}$$  
	and  $$p_\phi\rightarrow-i\frac{\partial}{\partial \phi}$$ .
	
	As a result one can get a two parameter family of WD equation as 
	\begin{eqnarray}
		\Big[\frac{A_2}{\phi}\frac{1}{a^l}\frac{\partial}{\partial a}\frac{1}{a^m}
		\frac{\partial}{\partial a}\frac{1}{a^n} +
		\frac{B_2\phi}{a^3}\frac{\partial^2}{\partial \phi^2}  
		+C_2\frac{1}{a^{2r}}
		\frac{\partial}{\partial a}\frac{1}{a^{2s}}\frac{\partial}{\partial 
			\phi}-V_2a^3\phi^{1-\frac{3}{k}}\Big]\psi(a,\phi)=0\label{eqm5}
	\end{eqnarray}	
	Here $(l,m,n)$, the triplet of real numbers, satisfy the condition $l+m+n=1$. But there are infinitely many possible choices for this triplet. So one can get infinite number of possible ordering. The commonly used choices are discussed below:
	
	\begin{itemize}
		\item D'Alembert Operator Ordering: $l=2$, $m=-1$, $n=0$.
		\item Vilenkin Operator Ordering: $l=0$, $m=1$, $n=0$.
		\item No Ordering: $l=1$, $m=0$, $n=0$.
	\end{itemize}
	
	The behaviour of the wave function can be affected by the factor ordering but the semi classical results remain unchanged due to the ordering problem. Now choosing the third option (i.e, no ordering), one can write the explicit form of WD equation as

	\begin{equation}
		\Big[\frac{A_2}{a\phi}\frac{\partial^2}{\partial 
			a^2} +\frac{B_2\phi}{a^3}\frac{\partial^2}{\partial \phi^2}+ \frac{C_2}{a^2}
		\frac{\partial^2}{\partial a\partial 
			\phi}-V_2a^3\phi^{1-\frac{3}{k}}\Big]\psi(a,\phi)=0\label{eqm6}
	\end{equation}
	Wave function of the Universe is nothing but the general solution of the above 
	2nd order hyperbolic partial differential equation. One can construct this solution by separating the Eigen functions of the WD operator as \cite{r38}
	\begin{equation}
		\psi(a,\phi)=\int{W(Q)\psi(a,\phi)dQ}\label{eqm7}
	\end{equation}       	
	where $Q$ is conserved charge, $W(Q)$ is weight function and $\psi$ is an eigen 
	function of the WD operator.  	
	In WD 
	operator, the minisuperspace variables $\{a,\phi\}$ are highly coupled so it is impossible to 
	have any explicit solution of WD equation even  with the separation of variable 
	method.
	Thus in the context of quantum cosmology, one may analyze this model using the new variables
	$(u,v)$ in the augmented space. 
	
	The canonically conjugate momenta can be written as
	\begin{eqnarray}
		p_u=\frac{\partial L}{\partial 
			\dot{u}}&=&e^{-2v}(2A_1\dot{u}+C_1\dot{v})=\mbox{Conseved}\nonumber\\
		p_v=\frac{\partial L}{\partial \dot{v}}&=&e^{-2v}(2B_1\dot{v}+C_1\dot{u})\label{eqm8}
	\end{eqnarray}
	Since in the transformed Lagrangian $u$ is cyclic then $p_u$ will be conserved.
	And hence the Hamiltonian of the system takes the form 
	\begin{equation}
		H=e^{2v}(A_3p_u^2+B_3p_v^2+C_3p_up_v)-V_3e^{-2v}\label{eqm9}	
	\end{equation}
	Here $A_3, B_3, C_3, V_3$ are arbitrary constants.
	
	Hence WD equation can be written as 
	\begin{equation}
		\Big[e^{2v}\Big(A_3\frac{\partial^2}{\partial u^2}+
		B_3\frac{\partial^2}{\partial v^2}+C_3\frac{\partial^2}{\partial u\partial 
			v}\Big)+V_3e^{-2v}\Big]\psi(u,v)=0\label{eqm10}
	\end{equation}  	
	The operator version of the conserved momentum takes the form 
	\begin{equation}
		i\frac{\partial_{\xi}(u,v)}{\partial u}=\Sigma_0\psi(u,v)\label{eqm11}
	\end{equation}  	
	Now we choose, $\psi(u,v)=A(u)B(v)$;
	
	Then from (\ref{eqm11}), we will get,
	\begin{eqnarray}
		i\frac{dA}{du}&=&\Sigma_0A\nonumber\\
		\implies A(u)&=&A_4e^{-i\Sigma_0u}\label{eqm12}
	\end{eqnarray}
	where $A_4$ is an integration constant.
	
	Now putting the expression of $A(u)$ in the WD equation (\ref{eqm10}), we will get a second order ordinary differential equation which is given by
	\begin{equation}
		B_3B''(v)-i\Sigma_0C_3B'(v)+(V_3e^{-4v}-A_3\Sigma_0)B(v)=0\label{eqm13}
	\end{equation}
	Solution of equation (\ref{eqm13}) is nothing but the non-oscillatory part of the wave function of the Universe for this model. The solution takes the form as
	\begin{figure}[h]
		\centering \includegraphics[height=5cm,width=7cm]{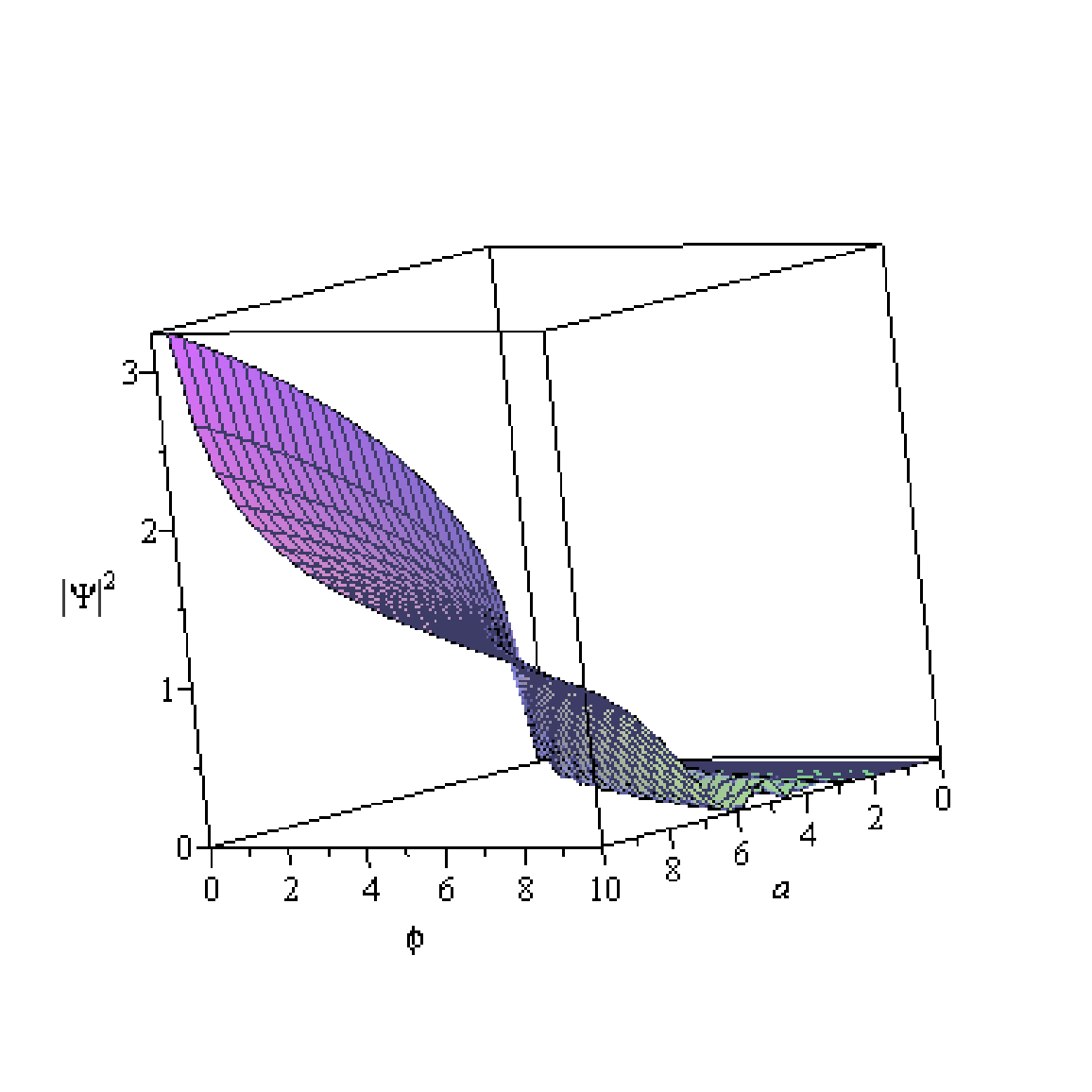}
		\caption{Wave function of the Universe}\label{f2}
	\end{figure}
	\begin{equation}
		B(v)=e^{ivA_5}\Big\{EJ_{(-B_4)}(C_4e^{-2v})+FJ_{(B_4)}(C_4e^{-2v})\Big\}\label{eqm14}
	\end{equation} 
	with $J$ as a Bessel function.\\ 	
	Therefore, the wave function of the Universe can be written as 
	\begin{equation}
		\psi(u,v)=e^{i(vA_5-u\Sigma_0)}\Big\{EJ_{(-B_4)}(C_4e^{-2v})+FJ_{(B_4)}(C_4e^{-2v})\Big\}\label{eqm15}
	\end{equation}
	The 3D figure in FIG.\ref{f2} shows the variation of the probability density (i.e., $|\psi|^2$) against the variables `$a$' and `$\phi$'. From the figure, it is clear that there is finite non--zero probability to have zero volume of the Universe.
	\section{BRIEF DISCUSSION AND CONCLUDING REMARKS}
	The paper investigates the homogeneous and isotropic Einstein-\ae ther scalar-tensor gravity model
	both for classical and quantum cosmology using Noether symmetry analysis. The Noether symmetry
	analysis is a powerful method which in some cases allows (i) to obtain exact dynamics and (ii) determining
	conserved Noether charge to be linked with some observable quantity. In the present work
	the above symmetry analysis shows two conserved quantity namely Noether charge and energy. Due
	to symmetry analysis it is possible to identify a cyclic co-ordinate (along the symmetry vector) so
	that the field equations become in a much simpler form and solutions are obtained. The cosmological
	solution of the present model shows the evolution of the universe from the early accelerated expansion to the present era of expansion as reflected in FIG.\ref{f1}.
	In the subsequent section (i.e., Section-{\bf VI}) quantum cosmology has been studied by forming the WD
	equation. The oscillatory part of wave function has been determined from the operator version of the
	conserved charge due to Noether symmetry. As a consequence, it is possible to have an explicit form
	of the wave function by solving the WD equation. The graph of probability density in FIG.\ref{f2} shows
	that quantum description can not avoid the initial singularity in the present cosmological model.


\frenchspacing
	\begin{thebibliography}{58}
		\bibitem{r11} T. Jacobson, {\it Phys. Rev. D} {\bf 81}, 10502 (2010); Erratum: {\it Phys. Rev. D} {\bf 82}, 129901 (2010).
		
		\bibitem{r12} I. Carruthers and T. Jacobson, {\it Phys. Rev. D} {\bf 83}, 024034 (2011).
		
		\bibitem{r13} D. Garfinkle and T. Jacobson, {\it Phys. Rev. Lett.} {\bf 107}, 191102 (2011).
		
		\bibitem{r14} W. Donnelly and T. Jacobson, {\it Phys. Rev. D} {\bf 82}, 064032 (2010).
		
		\bibitem{r1} A. Kostelecky, {\it Phys. Rev. D} {\bf 69}, 105009 (2004).
		
		\bibitem{r2} S.M. Carroll, H. Tam and I.K. Wehus, {\it Phys. Rev. D} {\bf 80}, 025020 (2009).
		
		
		\bibitem{r4} M. Mewes, {\it Phys. Rev. D} {\bf 99}, 104062 (2019).
		
		\bibitem{r5} S. M. Carroll and E. A. Lim,{\it  Phys. Rev. D} {\bf 70}, 123525 (2004)
		
		\bibitem{r6} C. Heinicke, P. Baekler and F.W. Hehl, {\it Phys. Rev. D} {\bf 72}, 025012 (2005)
		
		\bibitem{r7} C.A.G. Almeida, M.A. Anacleto, F.A. Brito, E. Passos and J.R.L. Santos, {\it Advances in High
			Energy Physics} {\bf 2017}, 5802352 (2017)
		
		\bibitem{r8} T. de Paula Netto, {\it Phys. Rev. D} {\bf 97}, 055048 (2018)
		
		\bibitem{r9} Z. Haghani, T. Harko, H.R. Sepangi and S. Shahidi,  {\it JCAP} {\bf 05}, 022 (2015).
		
		\bibitem{r10} T. Jacobson and D. Mattingly, {\it Phys. Rev. D} {\bf 70}, 024003 (2004).
		
		\bibitem{r14.1} C. Eling, T. Jacobson and M.C. Miller, {\it Phys. Rev. D.} {\bf 76}, 042003 (2007).
		
		\bibitem{r15} C. Bonvin, R. Durrer, P.G. Ferreira, G. Starkman and T.G. Zlosnik, {\it Phys. Rev. D} {\bf 77}, 024037
		(2008).
		
		\bibitem{r16} Y. Kukukakca and A.R. Akbarieh, {\it EPJC} {\bf 80}, 1019 (2020).
		
		\bibitem{r17} C.K. Ding, A.Z. Wang and X.W. Wang, {\it Phys. Rev. D} {\bf 92}, 084055 (2015).
		
		\bibitem{r18} R. Chan, M.F.A. da Silva and V.H. Satheeshkumar, {\it JCAP} {\bf 05}, 025 (2020).
		
		\bibitem{r19} K. Lin and Y.M. Wu, Mod. {\it Phys. Lett. A} {\bf 32}, 1750188 (2017).
		
		\bibitem{r20} R. Akhoury, D. Garnkle and N. Gupta, {\it Class. Quantum Grav.} 035006 (2018).
		
		\bibitem{r21} A. Coley and G. Leon, {\it Gen. Rel. Grav.} {\bf 51}, 115 (2019).
		
		\bibitem{r22} P. Horava, {\it Phys. Rev. D} {\bf 79}, 084008 (2009).
		
		\bibitem{r23} T.P. Sotiriou, M. Visser and S. Weinfurtner, {\it Phys. Rev. Lett.} {\bf 102}, 251601 (2009).
		
		\bibitem{r24} T.P. Sotiriou, M. Visser and S. Weinfurtner, {\it JHEP} {\bf 0910}, 033 (2009).
		
		\bibitem{r25} C.H. Brans and R.H Dicke, {\it Phys. Rev.} {\bf 124}, 925 (1965).
		
		\bibitem{r26} S. Sen and A.A. Sen, {\it Phys. Rev. D} {\bf 63}, 124006 (2001).
		
		\bibitem{r27} A. Bhadra, K. Sarkar, D.P. Natta and K.K. Nandi, {\it Mod. Phys. Lett. A} {\bf 22}, 367 (2007).
		
		\bibitem{r28} O. Bertolami and P.J. Martins, {\it Phys. Rev. D} {\bf 61}, 064007 (2000).
		
		\bibitem{r29} M. Tsamparlis, A. Paliathanasis, S. Basilakos and S. Capozziello, {\it Gen. Rel. Grav.} {\bf 45}, 2003 (2013).
		
		\bibitem{r30} M. Gasperini and G. Veneziano, Astropart. {\it Phys.} {\bf 1}, 317 (1993).
		
		{\bibitem{Bajardi:2022ypn}  F.Bajardi and  S.Capozziello, Noether Symmetries in Theories of Gravity {\it Cambridge University Press}, doi=10.1017/9781009208727 (2022).}
		
		{\bibitem{mu1} S. Capozziello, R. de Ritis, C. Rubano , P. Scudellaro, Riv.Nuovo Cim. 19, 1-114(1996).}
		
		{\bibitem{mu2} S. Capozziello, R. De Ritis, P. Scudellaro, Int.J.Mod.Phys.D 2, 
		373-379(1993),}
	
		{\bibitem{mu3}  S.Capozziello,  R. De Ritis, C. Rubano and  P. Scudellaro, Int.J.Mod.Phys.D 2,  463-476(1993)}
		
		{\bibitem{mu4}  S. Capozziello, R. De Ritis, P. Scudellaro, Int.J.Mod.Phys.D 3,  609-621 (1994).}
		
		{\bibitem{mu5} P. V. Moniz, Universe 8, 177(2022)}
		
		{\bibitem{mu6}  F. Bajardi, and S. Capozziello,  Int.J.Geom.Meth.Mod.Phys. 18 (2021) supp01, 2140002.}
		
	   {\bibitem{mu7} S. Capozziello, M. De Laurentis,  S.D.  Odintsov,  Eur.Phys.J.C 72,  2068(2012).}
	
	    {\bibitem{mu8} S. Capozziello,  G. Lambiase, Gen.Rel.Grav. 32,
	     673-696(2000).} 
    
       {\bibitem{mu9} S. Capozziello, M. De Laurentis, Int.J.Geom.Meth.Mod.Phys. 11,  1460004(2014).} 
    
    
		\bibitem{r31} A. Paliathanasis, G.Leon {\bf arXiv:2107.12546 [gr-qc]}.
		
		
		
		\bibitem{r32} S. Dutta, M. Lakshmanan and S. Chakraborty,  {\it Annals of Phys.} {\bf 407}, 1 (2019).
		
			
		
		
		\bibitem{r33} S. Dutta, M. Lakshmanan and S. Chakraborty,  {\it Phys. of Dark. Univ.} {\bf 32}, 100795 (2021).
		
		\bibitem{r34} M. Tsamparlis and A. Paliathanasis, {\it Class. Quant.Grav.} {\bf 29}, 015006 (2012).
		
		\bibitem{r35} S. Dutta and S. Chakraborty,  {\it Int. J. Mod. Phys. D} {\bf 25}, 1650051 (2016).
		
		
		\bibitem{r36} R. Bhaumik, S. Dutta and S. Chakraborty,  {\it EPJC} {\bf 82}, 1124 (2022).
		
		\bibitem{r37} S. Capozziello, A. Stabile and A. Troisi, {\it Class. Quant. Grav.} {\bf 24}, 2153 (2007).
		
		\bibitem{r38} Gravitation in astrophysics, Cargèse, 1986 / edited by B. Carter and J.B. Hartle.(New York : Plenum Press, 1986.)p-1-399.
		
		
		
		
	\end{thebibliography}
\end{document}